\documentstyle[aps,twocolumn,prl]{revtex}
\newcommand{\newc}{\newcommand}
\newc{\ra}{\rightarrow}
\newc{\lam}{\lambda}
\newc{\eps}{\epsilon}
\newc{\half}{\frac{1}{2}}
\newc{\gev}{\mbox{~GeV}}
\newc{\etal}{{\it et al.}\ }
\newc{\barr}{\begin{eqnarray}}
\newc{\earr}{\end{eqnarray}}
\newc{\beq}{\begin{equation}}
\newc{\eeq}{\end{equation}}
\newc{\eg}{{\it e.g.}\ }
\newc{\ie}{{\it i.e.}\ }
\newc{\kap}{\kappa}
\newc{\rpv}{$/\!\!\!\!R_p $}
\newc{\nonr}{\nonumber}
\newc{\eq}[1]{(\ref{eq:#1})}
\newc{\lab}[1]{\label{eq:#1}}
\newc{\Y}{{\bf Y}}
\newc{\ye}{{\Y}_E}
\newc{\yd}{{\Y}_D}
\newc{\yu}{{\Y}_U}
\newc{\lsim}{\stackrel{<}{\sim}}
\newc{\gsim}{\stackrel{>}{\sim}}
\newc{\er}[1]{\times \frac{m_{\tilde{#1}_R}}{100 \ {\rm GeV}}}
\newc{\ero}{\sqrt{m_{\tilde{\tau}}/100 \ {\rm GeV}}}
\newc{\ti}{\times}
\newc{\fm}{f(\tilde{m})}
\newc{\era}[1]{\sqrt{m_{\tilde{#1}}/100 \ {\rm GeV}}}
\newc{\erl}[1]{\times \frac{m_{\tilde{#1}_L}}{100 \ {\rm GeV}}}
\newc{\rnd}{(\frac{m_{\tilde{q}}}{\tilde{\Lambda} \ {\rm GeV}})^{5/2}}

\def\npb#1 #2 #3 #4 {Nucl.~Phys. B {\bf #1}, #2 (#3)#4 }
\def\plb#1 #2 #3 #4 {Phys.~Lett. B {\bf #1}, #2 (#3)#4 }
\def\prd#1 #2 #3 #4 {Phys.~Rev.  D {\bf #1}, #2 (#3)#4 }
\def\prl#1 #2 #3 #4 {Phys.~Rev.~Lett. {\bf #1}, #2 (#3)#4 }

\begin{document}
\draft
\twocolumn[\hsize\textwidth\columnwidth\hsize\csname @twocolumnfalse\endcsname
\hfill\parbox{8cm}{\raggedleft \today \\ DAMTP-1999-45 \\ hep-ph/9906209}\\

\title{Bounds on R-parity Violating Couplings at the Weak Scale and at
the GUT Scale}
\author{B.C. Allanach}
\address{DAMTP, Silver St, Cambridge,  CB3 9EW, UK}
\author{A. Dedes, H.K. Dreiner}
\address{Rutherford Appleton Laboratory, Chilton, Didcot, OX11 0QX, UK}

\maketitle

\begin{abstract}
We analyse bounds on trilinear R-parity violating couplings at the
unification scale by renormalising the weak scale bounds. We derive
unification scale upper bounds upon the couplings which are broadly
independent of the fermion mass texture assumed. The R-parity
violating couplings are factors of two to five more severely bounded
at the unification scale than at the electroweak scale. In the
presence of quark mixing, a few of the bounds are orders of magnitude
stronger than their weak scale counterparts due to new R-parity
violating operators being induced in the renormalisation between high
and low scales. These induced bounds are fermion mass texture
dependent. New bounds upon the weak scale couplings are obtained by
the requirement of perturbativity between the weak and unification
scales. A comprehensive set of the latest limits is included.
\end{abstract}

\vspace*{5mm}
]

\narrowtext

\section{Introduction}

When constructing the most general supersymmetric version of the
Standard Model (SM) there are baryon- and lepton-number violating
operators in the superpotential. These lead to rapid proton decay in
disagreement with the strict experimental bounds \cite{pdg}.
Therefore, an extra symmetry beyond the SM gauge symmetry, $G_{SM}=
SU(3)\times SU(2)\times U(1)$, must be imposed to protect the proton. 
In most cases the discrete multiplicative symmetry, R-parity ($R_p$)
\cite{farrar}, is chosen. This prohibits all baryon- and lepton-number
violating operators with mass dimension less or equal to four and
leads to the minimal set of couplings consistent with the data. The
resulting model is denoted the minimal supersymmetric standard model
(MSSM) \cite{mssm}.  However, the choice of $R_p$ is {\it ad
hoc}. There are other symmetries which are theoretically equally well
motivated \cite{cosmo} and which also prohibit rapid proton decay, \eg
both baryon-parity and lepton-parity. Baryon-parity even prohibits the
dangerous dimension five operators \cite{ibanez}.  For both baryon-
and lepton-parity, $R_p$ is violated (\rpv).

There is at present no direct experimental evidence for supersymmetry
and in particular no evidence for $R_p$ or \rpv \cite{pdg}. Theoretical 
models are our best guide. Ultimately we expect the weak-scale theory
to be embedded in a more fundamental unified theory formulated at a
significantly higher energy scale which should also be the origin of
$R_p$ or \rpv. There is an extensive list of models with $R_p$
\cite{kane}. However, \rpv\ grand unified models have been constructed
for the gauge groups $SU(5)$~\cite{hall,giudice,tamvakis,viss2,bere},
$SU(5)\times U(1)$~\cite{brahm,giudice,tamvakis}, $E_6$~\cite{rizzo}
 and $SO(10)$
\cite{giudice}, as well, and there are also string models of \rpv
\cite{bento}. At present no model is clearly preferred.

GUTs typically make predictions for ratios of Yukawa couplings,
\eg $m_b/m_\tau$ \cite{mbtau}. If the GUT is extended to include a
family symmetry for example via the Frogatt-Nielsen mechanism
\cite{froggatt} a prediction is obtained for the order of magnitude of
the Higgs Yukawa couplings. Since the quantum numbers are fixed, these
predictions can be extended to the \rpv-Yukawa couplings, see for
example \cite{chamseddine,nirx,lola}. In string theories the Yukawa
couplings are also in principle calculable.

When constructing an \rpv\ model at high energy, it is essential that
it is consistent with all experimental bounds on baryon- and
lepton-number violation. There are empirical bounds on all of the
\rpv-Yukawa couplings \cite{bgh,dreiner,bhat,group}, some of which are
quite strict. However, these bounds are all determined at the {\it
weak}-scale. They can therefore {\it not} be directly compared to the
predictions of the unified models, which are at the {\it GUT}-scale ($
M_{GUT}$) or higher. There are at present no bounds for \rpv-couplings
at the GUT scale. In order to compare the unification predictions with
the data we must employ the renormalization group equations (RGEs) for
the \rpv-Yukawa couplings. These equations have recently been given up
to two-loop order with the full \rpv\ flavour structure in
\cite{dedes}. The effect of running the couplings from the weak-scale
to the GUT scale can be substantial \cite{carlos,dedes}. 

It is the purpose of this letter to first update the weak-scale bounds
on \rpv-couplings and then to translate these bounds in a model
independent way into GUT scale bounds.\footnote{We do not discuss the
bounds on the bilinear coupling of the superpotential term $\kap_i L_i
H_2$, since this analysis needs knowledge of the $\mu$ parameter
possibly combined with the radiative electroweak symmetry breaking
scenario, and we postpone it to a forthcoming article.} For this we
employ the full one-loop RGEs of the \rpv-MSSM\footnote{Since we allow
for the fully generated flavour structure of the \rpv-couplings, a
full calculation at two-loop using the equations of \cite{dedes} would
be too complicated.} \cite{dedes}. In order to obtain the GUT-scale
bounds we assume a single coupling at the GUT scale in the current
eigenstate basis. After running the RGEs, we obtain a set of couplings
at the weak-scale, both from the flavour structure of the RGEs and
from the rotation into the mass eigenstate basis\footnote{For a
detailed discussion of the basis dependence of the \rpv-couplings see
\cite{davidson}.}. We compare this set with the existing weak-scale
bounds, including bounds on products of couplings. We also include
perturbativity bounds where they are more stringent than the empirical
ones. The bounds on the induced couplings often lead to significantly
stronger bounds on the GUT-scale couplings.

\section{Low Energy Bounds}
The first systematic study of low-energy bounds on the R-parity
violating Yukawa couplings was performed in \cite{bgh}. Since then
updates have been performed in \cite{bhat,dreiner}. More recently
there was a very nice thorough update of all the bounds on the
lepton-number violating couplings performed in \cite{gdr}. We present
in Table~\ref{table1} an updated version of only the best bound at two
standard deviations (2 sigma) on each coupling, respectively. For the
lepton-number violating couplings we update the results from
\cite{gdr} using the more recent data compiled by the particle data
group \cite{pdg}. The main difference to \cite{gdr} is due to the
improved data on the tau lepton parameters.  In the case of atomic
parity violation we have made use of new experimental data \cite{apv}
which is not yet included in \cite{pdg} and which leads to a new value
of $Q_W$. This differs from the Standard Model value by 2.5 sigma.
Thus we quote a three sigma bound. We do not include the recent bounds
obtained from $R_b$ \cite{yang}. Though they are the best bounds at 1
sigma, they are very weak at 2 sigma.

In Table~\ref{table2} we present a compilation of the bounds on the
product of two couplings. We have updated the bound from the decay
$K^+\ra\pi^+\nu{\bar\nu}$ \cite{Agashe} with the new data in
\cite{pdg,adler}.  We have then translated this bound into a bound on
the product of two couplings. In \cite{Agashe} the assumption was
explicitly made that at the weak scale there is only one dominant
coupling in the quark current basis. As described below this is not
necessarily true for our studies.

\section{Framework and Numerical Inputs}

The chiral superfields of the MSSM have the 
following $G_{SM}=SU(3)_c\otimes SU(2)_L\otimes U(1)_Y$ quantum numbers
\barr
L:(1,2,-\half),&&\quad {\bar E}:(1,1,1),\qquad\, Q:\,(3,2,\frac{1}{6}),
\nonr\\ 
{\bar D}:(3,1,-\frac{1}{3}),&&\quad
H_1:(1,2,-\half),\quad  H_2:\,(1,2,\half), \nonr \\
{\bar U}:(3,1,\frac{2}{3}). &&
\lab{fields}
\earr
We write the \rpv-MSSM superpotential as
\barr
W&=& \eps_{ab} \left[ (\ye)_{ij} L_i^a
H_1^b {\bar E}_j + (\yd)_{ij} Q_i^{ax} H_1^b {\bar D}_{jx} + \right. \nonr \\
&& 
(\yu)_{ij} Q_i^{ax} H_2^b {\bar U}_{jx} +
\frac{1}{2} \lam_{ijk} L_i^a L_j^b{\bar E}_k + \nonr \\ &&
\left. \lam'_{ijk} L_i^a Q_j^{xb} {\bar D}_{kx} +
\mu H_1^a H_2^b
+\kap^i L_i^a H_2^b \right]+ \nonr \\ &&
\frac{1}{2}\eps_{xyz} \lam''_{ijk} {\bar U}_i^x{\bar
D}_j^y{\bar D}^z_k. \label{SUPERP} 
\earr
We denote an $SU(3)$ colour index of the fundamental representation by
$x,y,z=1,2,3$. The $SU(2)_L$ fundamental representation indices are
denoted by $a,b,c=1,2$ and the generation indices by $i,j,k=1,2,3$.
We have introduced the three $3\times3$ matrices
\beq 
\ye,\quad \yd,\quad
\yu,\quad 
\lab{matrices} 
\eeq
for the $R_p$ conserving Yukawa couplings.

The boundary values of the running $\overline{DR}$ gauge couplings
$g_1(M_Z)$ and $g_2(M_Z)$ can be determined in terms of the
$\overline{MS}$ values of $\alpha^{-1}_{EM}(M_Z)=127.9$ and
$\sin^2\theta_W(M_Z)=0.2315$.  $M_{GUT}$ is found by the condition
$\alpha_1(M_{GUT})= \alpha_2(M_{GUT})$. Because above $M_Z$, we work
to one loop order only, $M_{GUT}=2.1\times10^{16}$~GeV is independent
of any Yukawa couplings.  The relation $\alpha_3(M_{GUT})=
\alpha_2(M_{GUT})$ is used to fix the strong coupling
constant\footnote{Note that the extracted value of $\alpha_3(M_Z)$ at
one-loop accuracy and without sparticle splitting threshold effects is
in excellent agreement with the experimental data. The R-Parity
violating couplings do not affect the running $\alpha_3$ at one loop
accuracy but they do at 2-loop level~\cite{dedes}.  However, the
effects are small ($\lsim$2\%
) for $\lam$, $\lam'$, $\lam'' \lsim
0.9$.}  $\alpha_3(M_Z)=0.118$.

We use the following experimentally determined fermion mass 
parameters\footnote{For quarks and leptons with masses less than 1 GeV, their
running masses have
been determined at the scale Q=1 GeV. As we go down to Q=1 GeV from
Q=$M_Z$ we decouple quarks or leptons when m(Q)=Q. In the case of the
top quark, only QCD corrections have been taken into the 
calculation of its running mass from the pole mass listed here.} 
(in GeV):
\barr
&&m_b(m_b) = 4.25 ,\ m_t^{\rm pole} = 175,\ m_\tau(m_\tau) = 1.777,
\label{inputs} \\
&&m_s =0.12 ,\ m_c(m_c) = 1.25 
,\ m_\mu = 0.105, \nonr\\
&&m_d = 0.006,\ m_u = 0.003,\ m_e = 0.000511,\nonr
\earr
where $m_i$ are listed in the $\overline{MS}$ renormalisation scheme
except for the pole mass of the top quark, $m_t^{\rm pole}$.  The
masses of the fermions are determined to 3 loops in QCD and 2 loops in
QED~\cite{arason} in the $\overline{MS}$ scheme and at the scale
$M_Z$. They are then converted into $\overline{DR}$ diagonal Yukawa
couplings using
\barr
h_{d,s,b,e,\mu,\tau}(M_Z) &=& \frac{m_{d,s,b,e,\mu,\tau}(M_Z)}{\sqrt{2}
v \cos 
\beta}, \nonr \\ 
h_{u,c,t}(M_Z) &=& \frac{m_{u,c,t}(M_Z)}{\sqrt{2} v \sin
\beta}, \label{fixmass}
\earr
where $v=246$~GeV is the standard model Higgs vacuum expectation value
(VEV), and $\tan \beta=v_2/v_1$ is the ratio of the two MSSM Higgs
VEVs. As an example study, throughout most of the paper we set $\tan
\beta=5$. We briefly discuss the case of $\tan\beta=35$ at the end.

We use central values of the mixing angles in the ``standard''
parametrisation of $V_{CKM}$ detailed in Ref.~\cite{pdg}
\barr
s_{12} = 0.2195,\ s_{23} =0.039 ,\ s_{13} = 0.0031.
\earr
We initially set the CP-violating phase $\delta_{13}=0$ but later we
examine $\delta_{13}=\pi/2$ to see the effects of CP violation.

For the purposes of the calculations we assume the entire MSSM
spectrum to be at the scale of the top quark mass, $m_t$, and
furthermore we assume a desert between $m_t$ and $M_{GUT}$.

\section{Numerical Procedure}
To obtain $\yu(M_Z)$ and $\yd(M_Z)$, assumptions have to be made about
the Yukawa matrices in the weak eigenbasis. To start with, we assume
that the mixing occurs only within the down quark sector, and that the
Yukawa matrices are Hermitian. We later also consider the other
extreme case where the mixing only occurs in the up quark sector. With
the definition of $\yd,\yu$ in Eq.\ref{SUPERP} and the mixing fully in
the down quark sector, we obtain
\beq
\yd(M_Z) = V_{CKM}^* {\yd}_{\mbox{diag}} (M_Z) V_{CKM}^T.\label{yd}
\eeq
${\yd}_{\mbox{diag}}(M_Z)$ is the diagonal matrix with $h_d(M_Z)$,
$h_s(M_Z)$, and $h_b(M_Z)$ along the diagonal. Thus $\yd(M_Z)$ is
determined uniquely in terms of its eigenvalues and the CKM matrix,
and all of the $R_p$-conserving couplings are defined at $\mu=M_Z$ in
the $\overline{DR}$ scheme.  Because the data on neutrino oscillations
are controversial, we do not include mixing of the charged leptons,
i.e. $\ye(M_Z)$ is set by its eigenvalues, the charged lepton masses
evaluated at $M_Z$.  

To begin, the system of all gauge couplings and all the Higgs Yukawa
couplings is evolved to $M_{GUT}$ using the one-loop RGEs of the
\rpv-MSSM \cite{carlos,dedes}. At the GUT scale, we then add only one
non-zero (and real) \rpv-coupling. This coupling is in the weak
current eigenbasis. All of the dimensionless couplings, now including
the \rpv-coupling, are then evolved down to $M_Z$. In the process more
than one non-zero \rpv-coupling is generated. The Higgs Yukawa
couplings evaluated at $M_Z$ in general lead to incorrect fermion
masses, so they are reset, as in Eqs.\ref{inputs},\ref{fixmass}. The
system of couplings is then re-evolved up to $M_{GUT}$ now including
the \rpv-couplings. At $M_{GUT}$, the \rpv-couplings can differ from
their initial values at $M_{GUT}$ and are reset. The process is
iterated until the system converges.

The \rpv-couplings thus obtained at the scale $M_Z$ are valid in the
weak eigenbasis. For comparison with experiment, the quark superfields
must be rotated to the quark mass eigenbasis. To do this, we follow
the procedure of Ref.~\cite{Agashe}. If we assume all the CKM mixing is
in the down quark sector only, we obtain the \rpv interactions
\barr
{\cal W}_{/\!\!\!\!R_p} &\supset& \lam'_{ijk} (V_{CKM}^\dagger)_{mk} \biggl [
N_i (V_{CKM})_{jl}D_l-E_iU_j \biggr ] \bar{D}_m \nonr \\
+& &\frac{1}{2}\lam''_{ijk}(V_{CKM}^\dagger)_{mj}
 (V_{CKM}^\dagger)_{nk} \bar{U}_i \bar{D}_m
\bar{D}_n.
\label{downmassbasis}
\earr
All superfields written in Eq.~\ref{downmassbasis} 
are in the quark mass eigenbasis,
contrary to those in Eq.~\ref{SUPERP}. The $\lam'$ terms have been expanded
into two SU(2) components containing $Q_i\equiv (U_i, D_i)$ and 
$L_i \equiv (N_i, E_i)$. 
Referring to Eq. \ref{downmassbasis}, we define the rotation
of the couplings to the quark mass basis (denoted with a tilde) 
\barr
\widetilde{\lam}'_{ijk} &=& \lam'_{ijm} (V_{CKM}^*)_{mk}, \label{pdcm}\\
\widetilde{\lam}''_{ijk} &=& \lam''_{imn}(V_{CKM}^*)_{mj}(V_{CKM}^*)_{nk}.
\label{ppdcm}
\earr
As shown in Ref.\cite{Agashe}, several \rpv interactions (as implied
by Eqs.~\ref{pdcm},\ref{ppdcm}) result in flavour changing neutral
currents (FCNC).  Upper bounds may then be obtained upon $\widetilde
{\lam}'$ and $\widetilde{\lam}''$ from FCNC data. Thus, starting with
a dominant \rpv-coupling in the weak eigenbasis at the GUT scale, we
evolve $\lam'_{ijk}, \lam''_{ijk}$ to the electroweak scale, causing
some of the \rpv-couplings to become non-zero through RG evolution.
At the electroweak scale, this system of \rpv-couplings is rotated
into the quark mass basis using Eqs.~\ref{pdcm},\ref{ppdcm}.

The resulting system of non-zero $\widetilde{\lam}'$ and $\widetilde
{\lam}''$ couplings valid at the electroweak scale is then checked
against the bounds summarised (together with their sources) in
Tables~\ref{table1},\ref{table2}.  Almost all of the bounds depend on
the sparticle masses.  The \rpv~GUT scale coupling is varied until the
couplings generated at $M_Z$ just pass the low-energy bounds.  The
value of the \rpv~GUT scale coupling at this point is then an upper
bound upon the non-zero R-parity violating GUT scale coupling. These
bounds are summarised in Table~\ref{table3}.

\section{Case Studies}

Here, we detail the results of the above procedure for various cases.
Initially, we present the bounds on GUT scale \rpv~couplings for a
simplified case in which there is no CP violating phase and zero
mixing, i.e.\ $V_{CKM}=1$. The results are displayed in
Tables~\ref{table1},\ref{table3}.  The perturbativity bounds upon
$\lam''_{ijk}$ presented in Table~\ref{table1} are in full agreement
with those given by Ref.\cite{goity}.  A two-loop calculation alters
the perturbativity bounds by up to 10\%~\cite{dedes}.  In this case,
there are no bounds caused by inducing new non-zero \rpv~couplings in
the renormalisation; a GUT scale bound is obtained by renormalising
the empirical bound on the dominant low energy coupling.  The explicit
dependence upon the sparticle masses in Table~\ref{table3} has been
demonstrated numerically. It is valid because of an approximate linear
relation between GUT and weak scale \rpv~couplings, valid in the limit
that they are small. This mass dependence is incorrect for cases where
the bound multiplied by a sparticle mass $\tilde{m}/100$ GeV is large
i.e. greater than 0.6. In those cases one can use the perturbativity
bound.  As can be seen from Table~\ref{table3}, bounds on
$\lam_{ijk}(M_{GUT})$ are approximately twice as severe than those on
$\lam_{ijk}(M_Z)$, whereas those on $\lam'_{ijk},\lam''_{ijk}
(M_{GUT})$ are three to five times as severe as their weak scale
counterparts.

Next, we examine the effects of quark mixing by assuming they it
occurs in the Hermitian $\yd$ given by Eq.\ref{yd}. Here, we set the
CP violating phase to zero. The results are displayed in
Table~\ref{table4} without parenthesis.  Obviously the bounds upon
$\lam_{ijk}(M_{GUT})$ in Table~\ref{table4} are identical to those in
Table~\ref{table3}, because the weak and mass bases of the leptons
have been assumed to be identical. When the bounds on
$\lam'_{ijk}(M_{GUT}), \lam''_{ijk}(M_{GUT})$ including quark mixing
effects are compared to those without mixing in Table~\ref{table3}, we
see a remarkable difference for many of the couplings. Many of them
are an order of magnitude more stringent when quark mixing has been
taken into account.  The $\lam''_{123}$ GUT scale couplings is
essentially unbounded in Table~\ref{table3} (or bounded by the limit
of perturbative believability), whereas in Table~\ref{table4}, the
bound becomes strengthened by an incredible seven orders of
magnitude. $\lam''_{113}$ becomes more constrained by a factor of
$500$.  For the $\lam'_{ijk}$ in Table~\ref{table3} that had the
strongest bound being that of perturbativity (for heavy sparticle
masses), down quark mixing effects imply that the empirical bounds are
the strongest.

In order to check the robustness of the bounds under changes in the assumed
R-parity conserving texture, we now perform the analagous analysis for the
case of mixing in a Hermitian $\yu$ only. For this case, Eq.\ref{yd} becomes
replaced by
\beq
\yu(M_Z) = V_{CKM}^T {\yu}_{\mbox{diag}} (M_Z) V_{CKM}^*,\label{yu}
\eeq
with $\yd(M_Z)={\yd}_{\mbox{diag}}$.
The superpotential terms in Eq.~\ref{downmassbasis} become
\barr
{\cal W}_{/\!\!\!\!R_p} &\supset& \lam'_{ijk}  \biggl [
N_i D_j-E_iU_l (V_{CKM}^\dagger)_{jl} \biggr ] \bar{D}_k \nonr \\
+& &\frac{1}{2}\lam''_{ijk}(V_{CKM})_{li}  \bar{U}_l \bar{D}_j
\bar{D}_k,
\label{upmassbasis}
\earr
for superfields in the quark mass eigenbasis.
This implies the rotation of \rpv-couplings
\barr
\widetilde{\lam}'_{ijk} &=& \lam'_{ilk} (V_{CKM}^*)_{jl} \label{pucm}\\
\widetilde{\lam}''_{ijk} &=& \lam''_{ljk}(V_{CKM})_{il} 
\label{ppucm},
\earr
supplanting Eqs.~\ref{pdcm},\ref{ppdcm}.
The rest of the numerical procedure is identical to that outlined in the
previous section.

Some of the bounds from mixing in the up quark sector (displayed in
square parenthesis in Table~\ref{table4}) are again remarkably
different to those without mixing in Table~\ref{table3}. There is
qualitatively less change in the $\lam'_{ijk}$ bounds from the
inclusion of up-quark mixing than down quark mixing, but some of the
$\lam''_{ijk}$ show an even larger strengthening effect. For example,
$\lam''_{212}$ instead of being bounded only by the perturbative
limit, acquires an empirical bound of 2.1$\times 10^{-9}$, obviously
very constraining upon relevant GUT models.  

To see the effect of CP violation, we pick $\delta_{13}=\pi/2$ as an
example and follow the above procedure for quark mixing in the down
quark sector (and subsequently in the up quark sector).  The bounds in
Table~\ref{table4} remain unchanged by the addition of CP-violation.
The GUT scale \rpv-couplings are taken to be real, but acquire small
phases from the renormalisation to the weak scale.  The largest
imaginary parts of couplings acquired occur when the dominant
couplings are large. The induced imaginary part of these couplings at
the weak scale is as large as $\sim 10^{-3}$ for quark mixing either
in the down quark or the up quark sector. For example, let us suppose we
start with the case where the mixing is in the down quark sector and
the dominant coupling at the GUT scale is $\lam''_{212}=-\lam''_{221}$
and is taken to be real. Then the renormalisation down to electroweak
scale induces non-zero and complex values for all of the other
$\lam''_{ijk}$.  The largest imaginary component is obtained for
$\lam''_{232}$ where Im$\lam''_{232}(M_Z)$=-Im$\lam''_{223}(M_Z)
\simeq 4\ti10^{-3}$.

To investigate how sensitive the GUT scale bounds are to the free
parameter $\tan \beta$, we performed another analysis with $\tan
\beta=35$ and no mixing.  For the case of the limits on $\lam_{ijk}$,
we find that the bound relaxes by up to 9\%.  In the cases of the
$\lam'_{ijk}$ or $\lam''_{ijk}$ \rpv-couplings we obtain a 30\% or 6\%
weakening of the the bound respectively.  Thus, to a 30\% accuracy
level, the bounds of the Table~\ref{table3} are stable over a large
range of $\tan \beta$.  Of course there is a strong dependence of the
perturbativity bounds in the regions $\tan\beta \lsim 3$ and
$\tan\beta \gsim 40$ upon the input value of
$\tan\beta$~\cite{dedes,roy4}.  The bounds from these values of $\tan
\beta$ are stronger than for $\tan \beta=5$ and so presenting the
bounds for $\tan \beta=5$ yields a conservative estimate.

\section{Summary}

We have examined changes in empirical bounds on \rpv-couplings as they
are renormalised to the unification scale, working to one loop
accuracy in perturbation theory but including all of the 45 trilinear
supersymmetric \rpv~ couplings.  The latest empirical bounds upon the
couplings have been collated in Tables~\ref{table1},\ref{table2}.  The
bounds upon $\lam'_{ijk}$ presented in Table~\ref{table1} in
parentheses are new except for $\lam'_{333}$, and are derived from the
requirement of perturbativity below the unification scale.  They are
the most stringent bounds on these couplings depending upon the squark
mass.  We have demonstrated that at high energy, the empirical bounds
upon the dominant \rpv~couplings are more severe than the empirical
bounds applied at $M_Z$ and are displayed in Table~\ref{table3}. The
bounds are made stronger by a factor of 2-5 from their
renormalisation.  These upper bounds are still applicable under
changes in the CP-violating phase and the inclusion of
quark-mixing. They are also approximately stable (at the 30\% level)
to changes in the parameter $\tan \beta$.  However, when quark mixing
is included some of the limits become several orders of magnitude more
severe than their weak scale counterparts due to new R-parity
violating operators being induced in the renormalisation between high
and low scales.  These very strong limits are dependent upon the
fermion mass texture, as we have demonstrated by calculating them in
the cases where the quark mixing is wholly within the down quark
sector, or wholly within the up quark sector.  While a CP violating
phase in the CKM matrix does not affect the bounds, the weak scale
\rpv~couplings acquire small imaginary components from the
renormalisation. The magnitudes of these phases are dependent upon the
mass texture assumed.  Since in general \rpv~terms can be induced via
non-renormalisable operators in GUT or other unified models this
analysis is hopefully useful for their phenomenology and construction.
A necessary condition upon any unified model is that it satisfy the
upper bounds given in Table~\ref{table3}.  Stronger constraints
arising from bounds upon induced couplings depend upon the fermion
mass texture assumed and so must be checked on a case-by-case basis.
The results presented here represent the most comprehensive collation
of bounds upon trilinear supersymmetric \rpv~couplings to date.

\begin{acknowledgements}
AD would like to thank A. Faraggi, H.M. Chan, and R. Roberts for
useful discussions and the theory group of Univ.\ of Minnesota for
their hospitality. This work was partially supported by PPARC\@. AD
acknowledges the financial support from the Marie Curie Research
Training Grant ERB-FMBI-CT98-3438.
\end{acknowledgements}

\twocolumn[\hsize\textwidth\columnwidth\hsize\csname @twocolumnfalse\endcsname

\begin{table*}
\caption{Latest $2\sigma$ limits on the magnitudes of weak scale 
trilinear R-parity violating couplings from indirect decays and
perturbativity. The dependence on the relevant superparticle mass 
is shown explicitly. When the perturbativity bounds are more stringent 
than the empirical bounds for masses  $m_{\tilde{l},\tilde{q}}$=1 TeV,
then we display them in parentheses. Where a bound without parentheses has no
explicit mass dependence shown, the mass dependence was too complicated to
detail here and a degenerate sparticle spectrum of 100~GeV is assumed.}
\label{table1}
\begin{tabular}{cccc}
ijk     & $\lam_{ijk}(M_Z)$\tablenote{Updated bounds from 
Ref.\cite{gdr,dreiner}. Bounds on $\lam_{121}$, $\lam_{122}$, $\lam_
{123}$ have been obtained from charged current universality \cite{bgh}. 
Bounds on $\lam_{131}$, $\lam_{132}$, $\lam_{231}$, $\lam_{232}$ and 
$\lam_{233}$ have been derived from~\cite{bgh} measurements of $R_{\tau}
=\Gamma(\tau\ra e\nu\bar{\nu})/\Gamma(\tau\ra \mu\nu
\bar{\nu})$ and $R_{\tau\mu}=\Gamma(\tau\ra\mu\nu\bar{\nu})/
\Gamma(\mu\ra e\nu\bar{\nu})$ \cite{pdg}. The bound on $\lam_{133}$ 
\cite{tata} has been obtained from the experimental limit on the 
electron neutrino mass \cite{pdg}.}
        & $\lam'_{ijk}(M_Z)$\tablenote{Bounds on $\lam'_{112}$, $\lam'_{113}$,
$\lam'_{121}$, $\lam'_{122}$, and $\lam'_{123}$ have been obtained
from charged current universality \cite{bgh}. The bound on $\lam'_{111}$ 
has been derived from neutrino-less double beta decay \cite{moha,beta}
where $\fm =(m_{\tilde{e}}/100\ {\rm GeV})^2\ti (m_{\tilde{\chi}^0}/
100\ {\rm GeV})^{1/2}$, and on $\lam'_{131}$ from atomic parity violation 
\cite{bgh,apv}. This latter bound is at the 3$\sigma$ level, 
since the data disagree with the standard model at the 2.5$\sigma$ level 
\cite{apv}. The bound on $\lam'_{132}$ comes from the forward-backward 
asymmetry in $e^+e^-$ collisions~\cite{bgh}. Bounds on $\lam'_{133}$,
$\lam'_{233}$ have been obtained from bounds on the neutrino masses 
\cite{tata} and on $\lam'_{211}$, $\lam'_{212}$, $\lam'_{213}$ from $R_\pi=
\Gamma(\pi\ra e\nu)/\Gamma(\pi\ra \mu\nu)$ \cite{bgh,gdr}. Bounds on 
$\lam'_{221}$, $\lam'_{231}$ come from $\nu_\mu$ deep inelastic scattering 
\cite{bgh,gdr} and on $\lam'_{222}$, $\lam'_{223}$ from the D-meson decays 
\cite{gdr}, $D\ra K l \nu$. The bounds without parentheses on $\lam'_{232}$, 
$\lam'_{331}$, $\lam'_{332}$, $\lam'_{333}$ have been derived from $R_l =
\Gamma(Z\ra {\rm had})/\Gamma(Z\ra l\bar{l})$ for $m_{\tilde{q}}=100$ 
GeV~\cite{ellis} and on $\lam'_{311}$, $\lam'_{312}$, $\lam'_{313}$ from 
$R_{\tau\pi}=\Gamma(\tau\ra \pi\nu_\tau)/\Gamma(\pi\ra \mu\nu_\mu)$ 
\cite{bgh,gdr}. The bounds on the couplings $\lam'_{321}$, 
$\lam'_{322}$ and $\lam'_{323}$ have been derived from $D_s$ decays 
\cite{gdr}, {\it i.e.,} $R_{D_s}=\Gamma(D_s\ra \tau \nu_\tau)/\Gamma
(D_s \ra \mu\nu_\mu)$. There are also bounds on $\lam'_{3j3}$ from $R_b$
\cite{yang} but these are weak at 2$\sigma$ level and thus not displayed.}
        & $\lam''_{ijk}(M_Z)$\tablenote{The indirect bounds on $\lam''_{ijk}$
existing in the literature are on $\lam''_{112}$ from double nucleon decay
\cite{goity} ($\tilde{\Lambda}$ is a hadronic scale and it can be varied
from 0.003 to 1 GeV and $\rnd$ from $2 \times 10^{11}$ to $10^{5}$ for 
$m_{\tilde{q}}=100$ GeV)   
and on $\lam''_{113}$ from neutron oscillations~\cite{zwirner,goity}
 for $m_{\tilde{q}}$=100 GeV. For $m_{\tilde{q}}$=200 (600)
GeV the bound on $\lam''_{113}$ is 0.002 (0.1). Bounds on $\lam''_{3jk}$
have been derived from
$R_l =\Gamma(Z\ra {\rm had})/\Gamma(Z\ra l\bar{l})$ at 
$1\sigma$ for $\tilde{m}=100$ GeV \cite{bat3} and, for heavy squark masses, is
not more stringent than 
the perturbativity bound, which is displayed in the parenthesis.}
\\ \tableline 
111     &    -            &$5.2\ti10^{-4}\ti\fm$&   -  \\
112     &    -            &$0.021 \er{s}$&                $10^{-15}\ti \rnd$ \\
113     &    -            &$0.021 \er{b}$&                  $10^{-4}$ \\
121     &$0.049 \er{e}$   &$0.043 \er{d}$&                $10^{-15}\ti \rnd$ \\
122     &$0.049 \er{\mu}$ &$0.043 \er{s}$&                 -\\
123     &$0.049 \er{\tau}$&$0.043 \er{b}$&        $(1.23)$\\
131     &$0.062 \er{e}$   &$0.019 \erl{t}$&        $10^{-4}$\\
132     &$0.062 \er{\mu}$ &$0.28 \erl{t}$ (1.04)&         $(1.23)$\\
133     &$0.0060\ero$     &$1.4\ti10^{-3}\era{b}$&  -\\
211     &$0.049 \er{e}$   &$0.059 \er{d}$&                 -\\
212     &$0.049 \er{\mu}$ &$0.059 \er{s}$&                 $(1.23)$\\
213     &$0.049 \er{\tau} $&$0.059 \er{b}$&       $(1.23)$\\
221     &   -             &$0.18\er{s}$ (1.12)&             $(1.23)$\\
222     &   -             &$0.21\er{s}$ (1.12)&             -\\
223     &   -             &$0.21\er{b}$ (1.12)&             $(1.23)$\\
231     &$0.070 \er{e}$   &$0.18\erl{b}$ (1.12)&            $(1.23)$\\
232     &$0.070 \er{\mu}$ &$0.56$ (1.04)&                  $(1.23)$\\
233     &$0.070 \er{\tau}$&$0.15\era{b}$&         -\\
311     &$0.062 \er{e}$   &$0.11\er{d}$ (1.12)&             -\\
312     &$0.062 \er{\mu}$ &$0.11\er{s}$ (1.12)&            $0.50$ (1.00)\\
313     &$0.0060\ero$     &$0.11\er{b}$ (1.12)&             $0.50$ (1.00)\\
321     &$0.070 \er{e}$   &$0.52\er{d}$ (1.12)&             $0.50$ (1.00)\\
322     &$0.070 \er{\mu}$ &$0.52\er{s}$ (1.12)&            -\\
323     &$0.070 \er{\tau}$&$0.52\er{b}$ (1.12)&           $0.50$ (1.00)\\
331     &  -              &$0.45$ (1.04)&           $0.50$ (1.00)\\
332     &  -              &$0.45$ (1.04)&           $0.50$ (1.00)\\
333     &  -              &$0.45$ (1.04)&          -      
\end{tabular}
\end{table*}
\vfill
]

\twocolumn[\hsize\textwidth\columnwidth\hsize\csname @twocolumnfalse\endcsname

\begin{table}
\caption{Current relevant upper limits on the values of products of
weak scale R-parity violating couplings for $\tilde{m}=100$ GeV.}
\label{table2}
\begin{tabular}{cc}
$|\lam_{1j1}\lam_{1j2}|$    &  $7\ti10^{-7}$~\tablenote{From 
$\mu\ra 3e$ \cite{roy3}}\\
$|\lam_{231}\lam_{131}|$    &  $7\ti10^{-7}$~\tablenote{From $\mu\ra 3e$
\cite{roy3}} \\
$|\lam_{231}\lam_{232}|$    & $5.3\ti10^{-6}$~\tablenote{From $\mu {\rm Ti} 
\rightarrow e {\rm Ti}$ at one loop \cite{santa}} \\
$|\lam_{232}\lam_{132}|$ & $8.4\ti10^{-6}$~\tablenote{From $\mu {\rm Ti} 
\rightarrow e {\rm Ti}$ at one loop \cite{santa}} \\
$|\lam_{233}\lam_{133}|$  & $1.7\ti10^{-5}$~\tablenote{From $\mu {\rm Ti} 
\rightarrow e {\rm Ti}$ at one loop \cite{santa}} \\
$|\lam_{122}\lam'_{211}|$ & $4.0 \ti10^{-8}$~\tablenote{From $\mu {\rm Ti} 
\rightarrow e {\rm Ti}$ at tree level \cite{santa}} \\
$|\lam_{132}\lam'_{311}|$ & $4.0 \ti10^{-8}$~\tablenote{From $\mu {\rm Ti} 
\rightarrow e {\rm Ti}$ at tree level \cite{santa}} \\
$|\lam_{121}\lam'_{111}|$ & $4.0 \ti10^{-8}$~\tablenote{From $\mu {\rm Ti} 
\rightarrow e {\rm Ti}$ at tree level \cite{santa}} \\
$|\lam_{231}\lam'_{311}|$ & $4.0 \ti10^{-8}$~\tablenote{From $\mu {\rm Ti} 
\rightarrow e {\rm Ti}$ at tree level \cite{santa}} \\
$|\lam'_{i1k}\lam'_{i2k}|$ & $2.2\ti10^{-5}$~\tablenote{From 
 $K\ra\pi\nu\bar{\nu}$ \cite{Agashe}. 
Also  
$|\lam'_{i11}\lam'_{i21}|\sim 10^{-6}$ from 
$\frac{\epsilon'}{\epsilon}$ \cite{abel} }\\
$|\lam'_{i12}\lam'_{i21}|$ & $10^{-9}$~\tablenote{From $\Delta m_K$ 
\cite{bere}}\\
Im$\lam'_{i12}\lam'^*_{i21}$ & $8\ti 10^{-12}$~\tablenote{From $\epsilon_K$
\cite{bere}}\\
$|\lam'_{113}\lam'_{131}|$ & $3\times 10^{-8}$~\tablenote{From $\Delta m_B$
\cite{babu}}\\
$|\lam'_{i13}\lam'_{i31}|$ & $8\times 10^{-8}$~\tablenote{From $\Delta m_B$
\cite{bere}}\\
$|\lam'_{1k1}\lam'_{2k2}|$ & $8\ti10^{-7}$~\tablenote{From $K_L\ra\mu e$
\cite{bere}}\\
$|\lam'_{1k1}\lam'_{2k1}|$ & $8.0\ti10^{-8}$~\tablenote{From $\mu{\rm
Ti}\ra 
e {\rm Ti}$ at tree level \cite{santa}}\\
$|\lam'_{11j}\lam'_{21j}|$ & $8.5\ti10^{-8}$~\tablenote{From $\mu{\rm
Ti}\ra 
e {\rm Ti}$ at tree level \cite{santa}}\\
$|\lam'_{22k}\lam'_{11k}|$ & $4\ti10^{-7}$~\tablenote{From $\mu{\rm
Ti}\ra 
e {\rm Ti}$ at tree level \cite{santa}}\\
$|\lam'_{21k}\lam'_{12k}|$ & $4.3\ti10^{-7}$~\tablenote{From $\mu{\rm
Ti}\ra 
e {\rm Ti}$ at tree level \cite{santa}}\\
$|\lam'_{22k}\lam'_{12k}|$ ($k$=2,3)& $2.1\ti10^{-6}$~\tablenote{From $\mu{\rm
Ti}\ra 
e {\rm Ti}$ at tree level \cite{santa}}\\
$|\lam'_{221}\lam'_{131}|$ & $2.0\ti10^{-6}$~\tablenote{From $\mu{\rm
Ti}\ra 
e {\rm Ti}$ at tree level \cite{santa}}\\
$|\lam'_{23k}\lam'_{11k}|$ & $2.1\ti10^{-6}$~\tablenote{From $\mu{\rm
Ti}\ra 
e {\rm Ti}$ at tree level \cite{santa}}\\
$|\lam'_{ij1}\lam'_{ij2}|$ (j $\neq$ 3) & $6.1\ti10^{-6}$~\tablenote{From $K$
and $B$ systems \cite{bat4}.}\\
$|\lam'_{i31}\lam'_{i32}|$  & $1.6\ti10^{-5}$~\tablenote{From $K$
and $B$ systems \cite{bat4}.}\\
$|\lam'_{i31}\lam'_{i12}|$  & $2.4\ti10^{-5}$~\tablenote{From $K$
and $B$ systems \cite{bat4}.}\\
$|\lam''_{i32}\lam''_{i21}|$ & 
$7.6\ti10^{-3}$~\tablenote{From non-leptonic 
decays of heavy quark mesons, $B^+\ra \bar{K}^0 + K^+$ \cite{roy}.}\\
$|\lam''_{i31}\lam''_{i21}|$ & 
$6.2\ti 10^{-3}$~\tablenote{From non-leptonic
decays of heavy quark mesons, $\Gamma(B^+\ra \bar{K}^0+\pi^+)/
\Gamma(B^+\ra J/\psi +K^+)$\cite{roy}.}\\
$|\lam''_{232}\lam''_{132}|$ & 
$2.5\ti10^{-3}$~\tablenote{From non-leptonic
decays of heavy quark mesons\cite{roy}.}\\
$|\lam''_{332}\lam''_{331}|$ & 
$4.8\ti10^{-4}$~\tablenote{From the contribution
of $K-\bar{K}$ mixing to the $K_L$-$K_S$ mass difference\cite{barbieri}.}
\end{tabular}
\end{table}
]

\twocolumn[\hsize\textwidth\columnwidth\hsize\csname @twocolumnfalse\endcsname

\begin{table*}
\caption{Bounds on the trilinear R-parity violating couplings at the GUT scale
which are in agreement with the low energy experimental bounds of
Tables~\ref{table1} and~\ref{table2}. The dependence of the
superparticle masses is shown explicitly, except where it is too complicated
and $\tilde{m}=100$~GeV is assumed. 
$**$ labels the perturbative
limit, for example 3.5. The input value of $\tan\beta$ has been chosen to
be $\tan\beta(M_Z)=5$.}
\label{table3}
\begin{tabular}{cccc}
ijk     & $|\lam_{ijk}(M_{GUT})|$  & $|\lam'_{ijk}(M_{GUT})|$   
& $|\lam''_{ijk}(M_{GUT})|$ \\ \tableline  
111     &    -          &$1.4\ti10^{-4}\ti\fm$&   -  \\
112     &    -          &$0.0059 \er{s}$& $2\ti10^{-16}\ti \rnd$ \\
113     &    -          &$0.0059 \er{b}$&                 
$2\ti10^{-5}$\tablenote{For $m_{\tilde{q}}$=200(600) GeV the bound is
$\lam''_{113}=\lam''_{131} \lsim 4\ti10^{-4}\, (3\ti10^{-2})$.} \\
121     &$0.032 \er{e}$ &$0.012 \er{d}$&        $2\ti10^{-16}\ti \rnd$ \\
122     &$0.032 \er{\mu}$&$0.012 \er{s}$&                 -\\
123     &$0.032 \er{\tau}$&$0.012 \er{b}$&        $(**)$\\
131     &$0.041 \er{e}$ &$0.0060 \erl{t}$&       $2\ti 10^{-5}$$^{\rm a}$\\
132     &$0.041 \er{\mu}$&$0.091 \erl{t}$ (1.65)\tablenote{From perturbativity
of the top Yukawa coupling.}&    $(**)$\\
133     &$0.0039\ero$   &$4.4\ti10^{-4}\era{b}$&  -\\
211     &$0.032 \er{e}$  &$0.016 \er{d}$&                 -\\
212     &$0.032 \er{\mu}$&$0.016 \er{s}$&                 $(**)$\\
213     &$0.032 \er{\tau}$&$0.016 \er{b}$&       $(**)$\\
221     &   -           &$0.051\er{s}$ ($**$)&            $(**)$\\
222     &   -           &$0.060 \er{s}$ ($**$)&           -\\
223     &   -           &$0.060\er{b}$ ($**$)&            $(**)$\\
231     &$0.046 \er{e}$ &$0.057\erl{b}$ ($**$)&           $(**)$\\
232     &$0.046 \er{\mu}$&$0.20$ (1.66)$^{\rm b}$&                $(**)$\\
233     &$0.046 \er{\tau}$&$0.048\era{b}$&        -\\
311     &$0.041 \er{e}$ &$0.031\er{d}$ ($**$)&            -\\
312     &$0.041 \er{\mu}$&$0.031\er{s}$ ($**$)&           $0.16$
(0.76)$^{\rm b}$ \\
313     &$0.0039\ero$   &$0.031\er{b}$ ($**$)&  $0.16$ (0.76)$^{\rm b}$\\
321     &$0.046 \er{e}$ &$0.17\er{d}$\tablenote{This bound can be used
only for small departures of sparticle masses from the electroweak scale.}
 ($**$)&                  $0.16$ (0.76)$^{\rm b}$\\
322     &$0.046 \er{\mu}$&$0.17\er{s}$$^{\rm c}$ ($**$)&                  -\\
323     &$0.046 \er{\tau}$&$0.17\er{b}$$^{\rm c}$ ($**$) &
                  $0.16$ (0.76)$^{\rm b}$\\
331     &  -            &$0.16$
 (1.66)$^{\rm b}$&                $0.16$ (0.76)$^{\rm b}$\\
332     &  -            &$0.16$ (1.66)$^{\rm b}$&       
          $0.16$ (0.76)$^{\rm b}$\\
333     &  -            &$0.16$ (1.66)$^{\rm b}$&       
         -      
\end{tabular}
\end{table*}
]

\twocolumn[\hsize\textwidth\columnwidth\hsize\csname @twocolumnfalse\endcsname

\begin{table*}
\caption{Basis dependent bounds
 on the trilinear R-parity violating couplings at the GUT scale
with the mixing assumed in the down[up] quark sector (see 
footnote in
Table.\ref{table1}). The value of 
 $\tilde{m}=100$~GeV for squarks and sleptons is assumed. 
 The input value of $\tan\beta$ and the hadronic scale $\tilde{\Lambda}$
have  been chosen to
be $\tan\beta(M_Z)=5$ and 300 MeV respectively.}
\label{table4}
\begin{tabular}{cccc}
ijk     & $|\lam_{ijk}(M_{GUT})|$  & $|\lam'_{ijk}(M_{GUT})|$   
& $|\lam''_{ijk}(M_{GUT})|$ \\ \tableline  
111     &    -          &$1.5\ti10^{-4}\;${\rm [}$1.5\ti10^{-4}${\rm ]}&  -  \\
112     &    -          &$6.7\ti10^{-4}\;${\rm [}$0.0059${\rm ]}&
  $4.1\ti 10^{-10}\;${\rm [}$4.1\ti 10^{-10}${\rm ]}\\
113     &    -          &$0.0059 \;${\rm [}$0.0059${\rm ]}&  $1.1\ti10^{-8}\;$
{\rm [}$2\ti10^{-5}${\rm ]} \\
121     &$0.032 $&$0.0015\;${\rm [}$6.7\ti10^{-4}${\rm ]}&$4.1\ti10^{-10}\;$
{\rm [}$4.1\ti 10^{-10}${\rm ]} \\
122     &$0.032 $&$0.0015\;${\rm [}$0.012${\rm ]}&                -\\
123     &$0.032 $&$0.012\;${\rm [}$0.012${\rm ]}&$1.3 \ti 10^{-7}\;$
{\rm [}$0.028${\rm ]}\\
131     &$0.041 $&$0.0027\;${\rm [}$0.0060${\rm ]}&      $1.1\ti 10^{-8}\;$
{\rm [}$2\ti10^{-5}${\rm ]}\\
132     &$0.041 $&$0.0027\;${\rm [}$0.091${\rm ]}&$1.3\ti 10^{-7}\;$
{\rm [}$0.028${\rm ]}\\
133     &$0.0039$&$4.4\ti10^{-4}\;${\rm [}$4.4\ti10^{-4}${\rm ]}&  -\\
211     &$0.032$  &$0.0015\;${\rm [}$0.016${\rm ]}&               -\\
212     &$0.032$  &$0.0015\;${\rm [}$0.016${\rm ]}&     $(**)\;$
{\rm [}$2.1\ti10^{-9}${\rm ]}      \\
213     &$0.032$  &$0.016\;${\rm [}$0.016${\rm ]}&      $(**)\;$
{\rm [}$1.0\ti10^{-4}${\rm ]}\\
221     &   -           &$0.0015\;${\rm [}$0.051${\rm ]}&       $(**)\;$
{\rm [}$2.1\ti10^{-9}${\rm ]}     \\
222     &   -           &$0.0015\;${\rm [}$0.060${\rm ]}&       -         \\
223     &   -           &$0.049\;${\rm [}$0.060${\rm ]}&  $(**)\;$
{\rm [}$0.028${\rm ]}  \\
231     &$0.046$        &$0.0027\;${\rm [}$0.057${\rm ]}& $(**)\;$
{\rm [}$1.0\ti10^{-4}${\rm ]}\\
232     &$0.046$        &$0.0028\;${\rm [}$0.20${\rm ]}& $(**)\;$        
{\rm [}$0.028${\rm ]} \\
233     &$0.046$        &$0.048\;${\rm [}$0.048${\rm ]}&           -       \\
311     &$0.041$        &$0.0015\;${\rm [}$0.031${\rm ]}&          -      \\
312     &$0.041$        &$0.0015\;${\rm [}$0.031${\rm ]}& $0.099\;$
{\rm [}$1.5\ti10^{-7}${\rm ]}  \\
313     &$0.0039$       &$0.0031\;${\rm [}$0.031${\rm ]}& $0.015\;$
{\rm [}$0.0075${\rm ]}   \\
321     &$0.046 $       &$0.0015\;${\rm [}$0.17${\rm ]}& $0.099\;$
{\rm [}$1.5\ti10^{-7}${\rm ]}   \\
322     &$0.046$        &$0.0015\;${\rm [}$0.17${\rm ]}&       -         \\
323     &$0.046$        &$0.049\;${\rm [}$0.17${\rm ]}&  $0.015\;$
{\rm [}$0.16${\rm ]}         \\
331     &  -            &$0.0027\;${\rm [}$0.16${\rm ]}& $0.015\;$
{\rm [}$0.0075${\rm ]}         \\
332     &  -            &$0.0028\;${\rm [}$0.16${\rm ]}& $0.015\;$
{\rm [}$0.16${\rm ]}      \\
333     &  -            &$0.091\;${\rm [}$0.16${\rm ]}& -
                
\end{tabular}
\end{table*}
]


\begin{references}
\bibitem{pdg} C. Caso et al, Eur. Phys. Jnl 3 (1998) 1.
\bibitem{farrar} G.R. Farrar, P. Fayet, Phys. Lett. B 76 (1978) 575.
\bibitem{mssm} For a recent extensive review see S.P. Martin,  
in G.L. Kane (ed.): Perspectives on supersymmetry, World Scientific (1998),
hep-ph/9709356.
\bibitem{cosmo} See the discussion in H. Dreiner, G.G. Ross
Nucl. Phys. B 410 (1993) 188, hep-ph/9207221, and the review 
in \cite{dreiner}.
\bibitem{ibanez} L.E. Ibanez, G.G. Ross, Nucl. Phys. B 368 (1992) 3.
\bibitem{kane} For a recent overview see the articles in G.L. Kane (ed.): 
Perspectives on supersymmetry, World Scientific (1998).
\bibitem{hall}L.J. Hall and M. Suzuki, \npb 231 419 1984 .
\bibitem{giudice} G.F. Giudice and R. Rattazzi, \plb 406 321 1997 .
\bibitem{tamvakis} K. Tamvakis, \plb 382 251 1996 ; \plb 383 307 1996 .
\bibitem{viss2} F. Vissani, Nucl. Phys. Proc. Suppl. {\bf 52A} 94 (1997) .
\bibitem{bere} R. Barbieri, A. Strumia and Z. Berezhiani, \plb 407 1997 250 .
\bibitem{brahm} D.E. Brahm and L.J. Hall, \prd 40 2449 1989 .
\bibitem{rizzo}T.G.~Rizzo, Phys. Rev. {\bf D46} (1992) 5102 .
\bibitem{bento} M.C. Bento, L. Hall, G.G. Ross, Nucl. Phys. B 292 (1987) 400.
\bibitem{mbtau} M. Chanowitz, J. Ellis, M.K. Gaillard, Nucl. Phys. B
128 (1977) 506.  
\bibitem{froggatt} C.D. Froggatt, H.B. Nielsen, Nucl. Phys. B (1979)
147.
\bibitem{chamseddine} A.H. Chamseddine, H. Dreiner, Nucl. Phys. B
(1996) 458, hep-ph/9504337.
\bibitem{nirx} T.~Banks, Y.~Grossman, E.~Nardi and Y.~Nir, Phys. Rev. 
{\bf D52} (1995) 5319, hep-ph/9505248; G.~Eyal and Y.~Nir, hep-ph/9904473.
\bibitem{lola} J. Ellis, S. Lola, G. G. Ross, Nucl. Phys. B 526 (1998) 115, 
hep-ph/9803308.
\bibitem{bgh} V. Barger, G.F. Giudice and T. Han, \prd 40 2987 1989 .
\bibitem{dreiner} H.K. Dreiner, published in {\it Perspectives on 
Supersymmetry}, ed. by G.L. Kane, World Scientific, 
{\tt hep-ph/9707435}.
\bibitem{bhat} G. Bhattacharyya, published in {\it Physics Beyond the
Standard Model; Beyond the Desert: Accelerator and Nonaccelerator
Approaches}, Tegernsee, Germany, June 1997, {\tt hep-ph/9709395}. 
\bibitem{group}R. Barbier, {\it et.al.,}, {\tt hep-ph/9810232}.
\bibitem{dedes}B.C. Allanach, A. Dedes and H.K. Dreiner, {\tt hep-ph/9902251}.
\bibitem{carlos} B. de Carlos and P. L. White, \prd 54 1996 3427.
\bibitem{davidson} J. Ellis and S. Davidson, Phys. Lett. B 390 (1997)
210, hep-ph/9609451; Phys. Rev. D 56 (1997) 4182, hep-ph/9702247;
S. Davidson, Phys. Lett. B 439 (1998) 63, hep-ph/9808425.
\bibitem{gdr} All bounds on $\lam$ and $\lam'$ have been updated by
F. Ledroit and G. Sajot, GDR-S-008 (ISN, Grenoble,1998). This can be
obtained at: http://qcd.th.${\rm u\!\!-\!\!psud}$.fr/GDR\_SUSY/
GDR\_SUSY\_PUBLIC/entete\_note\_publique. 
\bibitem{apv} S.C. Bennett, C.E. Wieman, Phys. Rev. Lett. 82 (1999)
2484, hep-ex/9903022; C.S. Wood, \etal, Science 275 (1997) 1759.
\bibitem{yang} J.M.~Yang, hep-ph/9905486.
\bibitem{Agashe} K. Agashe and M. Graesser, \prd 54 4445 1996 .
\bibitem{adler} E787 Collaboration (S. Adler et al.)
Phys. Rev. Lett. 79 (1997) 2204, hep-ex/9708031. 
\bibitem{arason}H. Arason, D.J. Castano, B. Keszthelyi,
 S. Mikaelian, E.J. Piard, P. Ramond, and  B.D. Wright,
\prd 46 3945 1992  and references therein.
\bibitem{goity} J.L. Goity and M. Sher, \plb 346 69 1995 ; {\it erratum}
\plb 385 500 1996 .
\bibitem{roy4}B. Brahmachari, P. Roy, \prd 50 39 1994 , {\it erratum}
{\bf ibid}, \prd 51 3974 1995 .
\bibitem{tata}R.M. Godbole, R.P. Roy and X. Tata, \npb 401 67 1993 .
\bibitem{moha} R.N. Mohapatra, Phys. Rev. D 34 (1986) 3457.
\bibitem{beta} M. Hirsch, H.V. Klapdor-Kleingrothaus and S.G. Kovalenko,
\prl 75 17 1995 ; \prd 53 1329 1996 .

\bibitem{ellis} J. Ellis, G. Bhattacharyya and K. Sridhar, Mod. Phys. Lett.
A {\bf 10} 1583 (1995).
\bibitem{zwirner}F. Zwirner, \plb 132 103 1983 .
\bibitem{bat3} G. Bhattacharyya, D. Choudhury and K. Sridhar,
\plb 355 193 1995 . Here we use the updated bounds as they are 
given by Ref.\cite{bhat}.
\bibitem{roy3}D. Choudhury and P. Roy, \plb 378 153 1996 .
\bibitem{santa}
K.~Huitu, J.~Maalampi, M.~Raidal and A.~Santamaria,
Phys. Lett. {\bf B430} (1998) 355
hep-ph/9712249.
\bibitem{abel} S.A.~Abel, Phys. Lett. {\bf B410} (1997) 173
hep-ph/9612272.
\bibitem{babu}K.S. Babu and R.N. Mohapatra, \prl 75 2276 1995 .
\bibitem{bat4} All the bounds of the combinations of $\lam'\ti\lam'$
couplings have been obtained recently by
 G. Bhattacharyya and A. Raychaudhuri, \prd 57 R3837 1998 . 
Here we display only bounds with $\lam'\ti\lam' \lsim 10^{-5}$ .
\bibitem{roy} C.E. Carlson, P. Roy and M. Sher, \plb 357  99 1995 .
\bibitem{barbieri} R. Barbieri and A. Masiero, \npb 267 679 1986 .
\end{references}
\end{document}